\magnification=\magstep1 
\hsize 32 pc
\vsize 42 pc
\baselineskip = 24 true pt
\def\cl{\centerline}
\def\vs {\vskip .4 true cm}
\cl {\bf A MODEL OF COMPOSITE ELECTRONS AND PHOTONS}
\vs
\cl {\bf T. PRADHAN}
\cl {Institute of Physics, Bhubaneswar-751005, INDIA}
\vs
\cl {PACS - 12.50 Ch. Models of quarks and leptons}
\vskip 0.8 true cm
\cl {\bf ABSTRACT}

It is shown that electrons and photons can be considered as composities of  particles 
belonging to the fundamental
representations of the extended Lorentz group 
$SU(3)\otimes SU(3)$ in (8+1) dimensional space-time which are held together by force
mediated by  gauge particles belonging to the regular representation of this group.
In this theory, spin states of electron,
positron and photon form a $8\otimes 8$ representation of the group.
One can also accommodate the electron neutrinos, W and Z bosons as well 
super partners of all these particles in this model. The theory allows
electron decay into photon and neutrino. 
There are only two parameters in this theory. These are the mass of the
fundamental particles and the fine structure constant of their coupling with gauge
particles. The former, estimated from the experimental limit on the electron life-time
works out to be greater than  $10^{-22}$ GeV and the latter works out to be unity
by making use of Dirac ground state energy formula for the composite neutrino which is
taken to be massless. These indicate that the compositeness of electron,
photon and other particles will be revealed at Planckian energies.

\vfill
\eject
Introduction of the concept of isospin space by Heisenberg [1] and symmetry of the
Hamiltonian with respect to rotation in this space provided a high degree of
order in the description of interaction among elementary particles. Extension of
this space and the symmetry by Gell-Mann and Ne'eman [2] revealed that all
hadrons are composed of quarks [3] held together by force provided by
coloured gluone [4]. Experimental evidence of this composite nature of
hadrons from inelastic electron scattering and jet events [5] is by now well
established. However, leptons are still regarded as elementary particles.
The highly successful Standard Model [6] of particle
interactions treats electrons as fundamental stable particles. However
experiments put a limit to its stability $(\tau > 2.7 \times 10^{23}$
years) [7]. This is an indication of a substructure of the electron with
composition provided by more fundamental objects. Quest for such objects have
been made through invocation of new internal symmetries [8]. Can
it be that this may be better achieved through extension of space-time symmetry similar to
Gell-Mann-Ne'eman extension of Heisenberg's internal symmetry which revealed
quark structure of hadrons ? It is the purpose of this communication to make an attempt
towards this end. To accomplish this we extend the Dirac equation of the (3+1)
dimensional space-time with invariance under the Lorentz group $SU(2)\otimes SU(2)$ to an (8+1)
dimensional space-time with invariance under the extended Lorentz group 
$SU(3)\otimes SU(3)$.

The Dirac equation in spinor representation [9]
$$\eqalign{ (p_0-\vec\sigma .\vec p)\xi_p & = m\eta_p\cr
(p_0+\vec\sigma .\vec p)\eta_p & = m\xi_p\cr}\eqno{(1)}$$
where 
$$\xi_p =  \bigg ( \matrix{\xi_1\cr \xi_2\cr} \bigg), \ \eta_p = \bigg  (\matrix{\eta_1\cr\eta_2\cr}\bigg )\eqno{(2)}$$
are form-invariant under the Lorentz group of transformations 
$$\eqalign{\delta p_i &  = \varepsilon_{ijk} p_j\theta_k - p_0\phi_i\cr
\delta p_0 & = - p_i\phi_i\cr i,j & = 1,2,3\cr}\eqno{(3)}$$
under which
$$\eqalign{\delta\xi_p & = i {\sigma_i\over 2} (\theta_i-i\phi_i)\xi_p  = i[K_i,\xi_p]
(\theta_i-i\phi_i)\cr 
\delta\eta_p & = i {\sigma_i\over 2} (\theta_i+i\phi_i)\eta_p  = i[L_i,\eta_p]
(\theta_i+i\phi_i)\cr }\eqno{(4)}$$
where the generators
$$\eqalign{K_i & = {1\over 2}\xi_p^+\sigma_i\xi_p\cr 
L_i & = {1\over 2}\eta_p^+\sigma_i\eta_p\cr }\eqno{(5)}$$
of the Lorentz group obey $SU_2(\xi)\otimes SU_2(\eta)$ algebra
$$\eqalign{[K_i,K_j] & = i\varepsilon_{ijk} K_k\cr
[L_i,L_j] & = i\varepsilon_{ijk} L_k\cr
[K_i,L_j] & = 0\cr i,j,k & = 1,2,3\cr}\eqno{(6)}$$

We shall now extend this algebra (6) to $SU_3(\xi)\otimes SU_3(\eta)$:
$$\eqalign{[K_a,K_b] & = if_{abc} K_c\cr
[L_a,L_b] & = if_{abc} L_c\cr
[K_a,L_b] & = 0\cr
a,b,c  & = 1,2,3......8\cr}\eqno{(7)}$$
and extend equations (1) to
$$\eqalign{(p_0 - {1\over 2}\lambda_a p_a)\xi & = m\eta\cr
(p_0 + {1\over 2}\lambda_a p_a)\eta & = m\xi\cr}\eqno{(8)}$$
where
$$\xi =\bigg (\matrix{\xi_1\cr\xi_2\cr\xi_2\cr}\bigg )  \  \eta = \bigg (
\matrix{\eta_1\cr\eta_2\cr\eta_3\cr}\bigg )\eqno{(9)}$$
$$K_a = {1\over 2}\xi^+\lambda_a\xi ,\ \ L_a = {1\over 2}
\eta^+\lambda_a\eta\eqno{(10)}$$
$\lambda_a$ being $3\times 3$ SU(3) matrices. It can be verified that the pair
equations (8) are equivalent to three pairs of equations one of which is
equation (1) and the rest two are
$$\eqalign{(p_0-\vec\sigma.\vec\pi)\xi_{\pi} & = m\eta_{\pi}, \qquad 
(p_0-\vec\sigma.\vec\kappa)\xi_{\kappa}  = m\eta_{\kappa} \cr
(p_0+\vec\sigma.\vec\pi)\eta_{\pi} & = m\xi_{\pi},  \qquad
(p_0+\vec\sigma.\vec\kappa)\eta_{\kappa}  = m\xi_{\kappa}\cr}\eqno{(11)}$$
where
$$\xi_{\pi} = \bigg (\matrix{\xi_1\cr \xi_3}\bigg ), \ \eta_{\pi} = \bigg (\matrix{\eta_1\cr\eta_3\cr} \bigg ), \
\xi_{\kappa} = \bigg (\matrix{\xi_2\cr\xi_3\cr}\bigg ),  \ \eta_{\kappa} = \bigg (\matrix{\eta_2\cr\eta_3\cr}\bigg )\eqno{(12)}$$
and
$$\vec\pi = (p_4,p_5,{p_3\over\sqrt 3}) \ \ \vec\kappa = (p_6,p_7,{p_8\over\sqrt 3})\eqno{(13)}$$
The three pairs  equations appearing in (1) and (11)
represent the dynamics of fundamental building
blocks of our model. Noting that 
$K_8 = {1\over 2} \xi^+ \lambda_8 \xi = {1\over\sqrt 3} (\xi^+_p\xi_p-{1\over 2}
\xi^+_{\pi}\xi_{\pi}-{1\over 2}\xi^+_{\kappa}\xi_{\kappa})$,
their weight diagram will be as given in Fig.1.
\vskip  2.0 true cm
\hskip 2.0 true cm Fig.1

Our next task is to constructed electrons positrons and photons as well as
neutrino, Z and W bosons using the fundamental particles discussed above. In this
connection we note that since under parity transformation $\xi\leftrightarrow \eta$ it is
adequate to construct composites in $\xi$-space only. In this space the
electron and the positron can be constructed from the $SU_p(2)$ doublets $\xi^{(\mu)}_p,
\xi^{(d)}_p, \xi_{\pi}$ and $\xi_{\kappa}$ of this group. The superscripts u and d
stand for spin up and down.
$$\eqalign{e^{(u)} & = {1\over \sqrt 2} \xi_p^{(u)} (\xi_{\pi}^+ +\xi^+_{\kappa}) \cr
e^{(d)} & = {1\over \sqrt 2} \xi_p^{(d)} (\xi_{\pi}^+ +\xi^+_{\kappa}) \cr
\overline e^{(u)} & = {1\over \sqrt 2} \xi_p^{+(u)} (\xi_{\pi} +\xi_{\kappa}) \cr
\overline e^{(u)} & = {1\over \sqrt 2} \xi_p^{+(d)} (\xi_{\pi} +\xi_{\kappa}) \cr}
\eqno{(14)}$$
The right and left circularly polarized photon states can be
constructed from $\xi_p $ and $\xi^+_p$:
$$\eqalign{A^{(R)} & = \xi^{(u)}_p \xi^{+^(d)}_p\cr
A^{(L)} & = \xi^{(d)}_p \xi^{+^(u)}_p\cr}\eqno{(15)}$$
The states 
$$ \eqalign{ A^{(3)}& = {1\over\sqrt 2} (\xi_p^{(u)}\xi_p^{+(u)} -  \xi_p^{(d)}\xi_p^{+(d)})\cr
 A^{(8)} &  = {1\over\sqrt 6} (\xi_p^{(u)}\xi_p^{+(u)} +  \xi_p^{(d)}\xi_p^{+(d)}-2\xi_{\pi}
\xi^+_{\pi})\cr}\eqno{(16)}$$
do not bind as the gauge coupling is repulsive for antiparallel spins [10].
The composites given in (14), (15) and (16) from an octet and are represented in the
weight diagram given in Fig.2.
\vskip 2.0 true cm
\centerline {\bf Fig.2}
\vs
The neutrinos states are obtained by replacing $(\xi_{\pi}+\xi_{\kappa}){1\over\sqrt 2}$
in (14) by ${1\over\sqrt 2} (\xi_{\pi}-\xi_{\kappa})$. For accomodating Z and
W bosons it would be necessary modify the representation (15) and (16)
for photon states to: 
$$ A = (\xi_p\xi^+_p)(\xi^+_{\pi}\xi_{\pi}) = \Psi_p\Phi_p\eqno{(17)}$$
where $\Psi_p =\xi_p\xi^+_p$ is a second rank spinor and $\phi_{\pi} =
\xi^+_{\pi}\xi_{\pi}$ is a scalar in the $(p_0, p_1, p_2, p_3)$ space. In that case
$$\eqalign{ Z & = \Psi_p\Phi_{\kappa}\cr
W & = \Psi_p\Phi_{\kappa\pi}\cr}\eqno{(18)}$$
where $\Phi_{\kappa} =\xi^+_{\kappa}\xi_{\kappa}$ and 
$\Phi_{\kappa,\i} = \xi^+_{\kappa}\xi_{\pi}$ are scalars. It is to be
noted that there exist composite scalar particles.
$$\eqalign{ \tilde e & = {\phi_p\over\sqrt 2}(\xi^+_{\pi}+\xi^+_{\kappa})\cr
 \tilde\nu & = {1\over\sqrt 2}\phi_p(\xi^+_{\pi}-\xi^+_{\kappa})\cr}\eqno{(19)}$$ 
(where $\phi_p =\xi^+_p\xi_p$ is a scalar) which can be taken as super-partners of
electron and neutrino. Similarly there exist spin ${1\over 2}$ composites
$$\eqalign{\tilde A & = \xi_p\phi^+_{\pi}\cr
\tilde Z & = \xi_p\phi^+_{\kappa}\cr
\tilde W & = \xi_p\phi^+_{\kappa\pi}\cr}\eqno{(20)}$$
which can be taken as super partners of A, Z and W.

The gauge coupling of the fundamental particles is obtained from local version
of eqns.(1), (8) and (11) which are obtained by the replacements
$$ p_0\longrightarrow (p_0-gG_0 ) \qquad \vec p \longrightarrow (\vec p - g\vec G_p)$$
$$ \vec\pi \longrightarrow
(\vec\pi - g\vec G_{\pi} ) \qquad \vec\kappa \longrightarrow (\vec\kappa-g\vec G_k)\eqno{(21)}$$
in the respective equations. The interaction Lagrangian obtained from the
local equations work out to be
$$\eqalign{{\cal L}_{int} & = g\overline\psi_p\gamma_{\mu}
(v^{(p)}_{\mu}+\gamma_5 a_{\mu}^{(p)})\psi_p \cr & \ 
+g\overline\psi_{\pi}\gamma_{\mu}(v^{(\pi)}_{\mu}+\gamma_5 a_{\mu}^{(\pi)})\psi_{\pi} \cr & \ 
+g\overline\psi_{\kappa}\gamma_{\mu}(v^{(\kappa)}_{\mu}+\gamma_5 a_{\mu}^{(\kappa)})\psi_{\kappa} \cr}\eqno{(22)}$$
where                                           
$$\psi_p =\bigg (\matrix{\xi_p\cr \eta_p\cr}\bigg ) \ \psi_{\pi} = \bigg (
\matrix{\xi_{\pi}\cr\eta_{\pi}}\bigg ) \ \psi_{\kappa} = \bigg (\matrix{\xi_{\kappa}\cr\eta_{\kappa}}\bigg )\eqno{(23a)}$$
$$\eqalign{v^{(p)}_{\mu} & =  {1\over 2} [ G^{(p)}_{\mu}(\xi)+G_{\mu}^{(p)} (\eta) ]\cr
a^{(p)}_{\mu} & =  {1\over 2} [ G^{(p)}_{\mu}(\eta)-G_{\mu}^{(p)} (\xi) ]\cr}\eqno{(23b)}$$
and similar definitions for $v_{\mu}^{(\pi)}, a_{\mu}^{(\pi)} ,v^{(\kappa)}_{\mu}$ 
and $a^{(\kappa)}_{\mu}$ with $\mu = 0,4,5,8 $ for $\pi$ fields and $\mu = 0,6,7,8$ for
$\kappa$ fields.

Equation of motion for the gauge fields
$$\eqalign{ f_p & = {1\over 2} \sigma_i (e_i+ib_i) \cr
a & = 1,2,3,.....8 \cr} \eqno{(24)}$$
where $e_i$ and $b_i$ are components of the field tensor
$$F_{\mu\nu} = \partial_{\mu} v_{\nu}-\partial_{\nu} v_{\mu}+\varepsilon_{\mu\nu\lambda\xi}\partial_{\lambda}
a_{\xi}\eqno{(25)}$$
can be written as [11]
$$(p_0-\vec\sigma.\vec p)f_p = 0\eqno{(26a)}$$
In the extended space this can be written as
$$(p_0 - {1\over 2} \lambda_a p_a) f = 0\eqno{(26b)}$$
with $f = {1\over 2} \lambda_a (e_a+ib_a) = \lambda_a f_a$. This equation is
equivalent to the following six equations (27) and (29) if we set 
$f_3 = f_8 = 0$.
$$\eqalign{(p_0-\vec\sigma\cdot\vec p) f_p &= 0 \cr
(p_0-\vec\sigma\cdot\vec \pi) f_{\pi} &= 0 \cr
(p_0-\vec\sigma\cdot\vec \kappa) f_{\kappa} &= 0  \cr}\eqno{(27)}$$
where
$$ f_p = {1\over 2}\vec\sigma\cdot\vec f_p \qquad f_{\pi}={1\over 2}\vec\sigma\cdot\vec f_{\pi} 
\qquad f_{\kappa} = {1\over 2}\vec\sigma\cdot\vec f_{\kappa}\eqno{(28a)}$$
with
$$\vec f_p = (f_1,f_2,f_3) \qquad f_{\pi} = (f_4,f_5,{f_8\over \sqrt 3}) \qquad \vec f_{\kappa} =
(f_6,f_7,{f_8\over\sqrt 3})\eqno{(28b)}$$
and 
$$\eqalign{ (p_0-\vec\sigma\cdot \vec p)h_p = 0 \cr
 (p_0-\vec\sigma\cdot\vec\pi)h_{\pi} = 0 \cr
 (p_0-\vec\sigma.\cdot\vec\kappa)h_{\kappa} = 0 \cr}\eqno{(29)}$$
where
$$h_p =\bigg (\matrix {f_{\pi}^{(d)}\cr f^{(d)}_{\kappa}\cr}\bigg ) \ \
h_{\pi} = \bigg (\matrix {f_p^{(d)}\cr f^{(u)}_{\kappa}\cr}\bigg ) \ \
h_{\kappa} = \bigg (\matrix {f_p^{(u)}\cr f^{(u)}_{\pi}\cr}\bigg ) \eqno{(30a)}$$
with
$$\eqalign{ f^{(u)}_p & = (f_1 + if_2) \qquad f_{\pi}^{(u)} = (f_4 + i f_5) \qquad f^{(u)}_{\kappa} = 
(f_6 + i f_7)\cr
f^{(d)}_p & = (f_1 - if_2) \qquad f_{\pi}^{(d)} = (f_4 - i f_5) \qquad f^{(d)}_{\kappa} = 
(f_6 - i f_7)\cr}\eqno{(30b)}$$
It will be noted that while equations (27) describe dynamics of $p, \pi$ and $\kappa$ space
triplet (bosonic) gauge fields equations (29) describe the dynamics of the
corresponding doublet (fermionic) gauge fields. One of these  three octats is 
represented in the weight diagram given in Fig.3.
\hskip 4.0 true cm Fig.3
\vskip 1.0 true cm
Local versions of eqns(27) and (29) can be obtained by adopting prescription (21).

We shall now show how the gauge coupling constant of the fundamental particles can
be determined from the masslessness of the neutrino and their mass can be
estimated from the experimental limit on the life-time of the electron.
The ground state energy of the composite neutrino, as given by Dirac theory,
is 
$${\cal E} = m \sqrt{1-\alpha^2_g},\eqno{(31)}$$
Setting this to zero gives
$$\alpha_g = 1\eqno{(32)}$$
In our model, the composite electron can make a transition to the neutrino
state by emitting a M1 photon : 
$$e\longrightarrow \nu+ \gamma\eqno{(33)}$$
Since electric charge is not a fundamental quantum number in our model, there is
no problem with non-conservation of charge in this decay. The life-time of
this transition is given by
$${1\over\tau} = {4w^3\over 3} < \nu \mid M\mid e>\mid^2\eqno{(34)}$$
where M is the transition magnetic dipole operator and $w$ is the frequency of the emitted
photon which, in this case equals electron mass. From dimensional considerations
$$<\nu \mid M \mid e > \sim \alpha_g/m\eqno{(35)}$$
Substituting this in eqn(34), we get
$$m^2 = \alpha_g^2 m^3_e\tau\eqno{(36)}$$
Using the experimental limit 
$$\tau > 2.7\times 10^{23} years \eqno{(37)}$$
on the electron life-time, we get
$$m > 10^{22} GeV \eqno{(38)}$$
This corresponds to distances less than $10^{-36} cm$ which is in the
Planckian regime. This is an indication of the fact that the 
compositeness of the electron, photon and other particles discussed in the
note will be revealed at Planckian energies.

In order to accommodate other leptons and quarks it would be necessary to extend the
Lorentz group to $SU(5)\otimes SU(5)$. Details of this model will be presented
in a separate communication,
\vfill
\eject
\centerline {\bf References}
\vskip .3 true pc
\item {1.} W. Heisenberg, Z Physik {\bf 77} 1 (1932).
\item {2.} M. Gell-Mann, Cal. Inst. Tech. Rep. CTSL-20, Pasadena Calif 91961),
Phys. Rev. {\bf 125} 1067 (1962)
\item {} Y. Ne'eman, Nucl. Phys. {\bf 26} 222 (1961).
\item {3.} M. Gell-Mann, Phys. Lett. {\bf 8} 214 (1964)
\item {} G. Zweig (unpublished CERN report).
\item {4.} O.W. Greenberg, Phys. Rev. Lett. {\bf 13} 598 (1964).
\item {} M. Gell-Mann, Acta. Phys. Austriacca Supp. {\bf 9} 733 (1972).
\item {5.} R.P. Feynman ``Photon - Hadron Interactions'' (Reading, Mass. Benjamin 1972).
\item {} G. Hansen et. al. Phys. Rev. Lett. {\bf 35} 1609 (1975).
\item {6.} S.L. Glashow, Nucl. Phys. {\bf 22} 579 (1961).
\item {} S. Weinberg, Phys. Rev. Lett. {\bf 19} 1264 (1967)
\item {} A. Salam in ``Elementary Particle Theory'' ed. N. Svartholm
(Almquist and Wicksell, Stockholm 1968).
\item {7.} M. Aguilar-Benitez et. al. Phys. Rev. {\bf D50}, 1173 (1994).
\item {8.} For a review of preon models see M.E. Peskin in Proceedings of
International Symposium on Lepton and Photon Interactions at High
Energies ``Bonn 1991 p.880.
\item {9.} See for instance ``Quantum Electrodynamics'' by V.B. Beretetskii, 
E.M. Lifsitz and L.P. Pitaevskii (Pergmann Press, second edition, section 20).
\item {10.} T. Pradhan, R.P. Malik and P.C. Naik, Pramana {\bf 24} 77 (1985).
\item {11.} O. Leporte and GE. Uhlenbeck, Phys. Rev. {\bf 37} 1380 (1931).
\vfill
\eject
\centerline {\bf Figure Caption}
\vskip .3 true cm
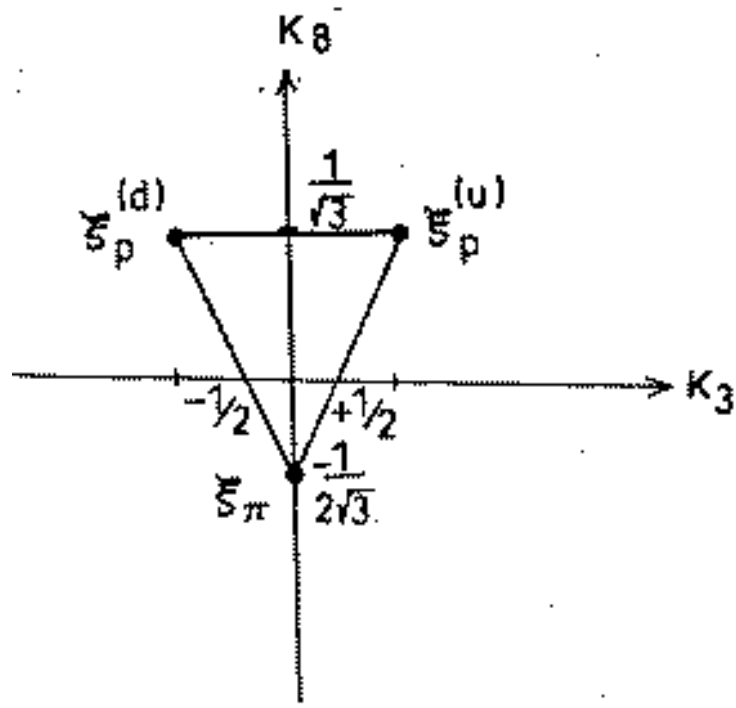
\item {Fig.1.} Weight diagram for the fundamental particles.
\item {Fig.2.} Weight diagram for electron and photon
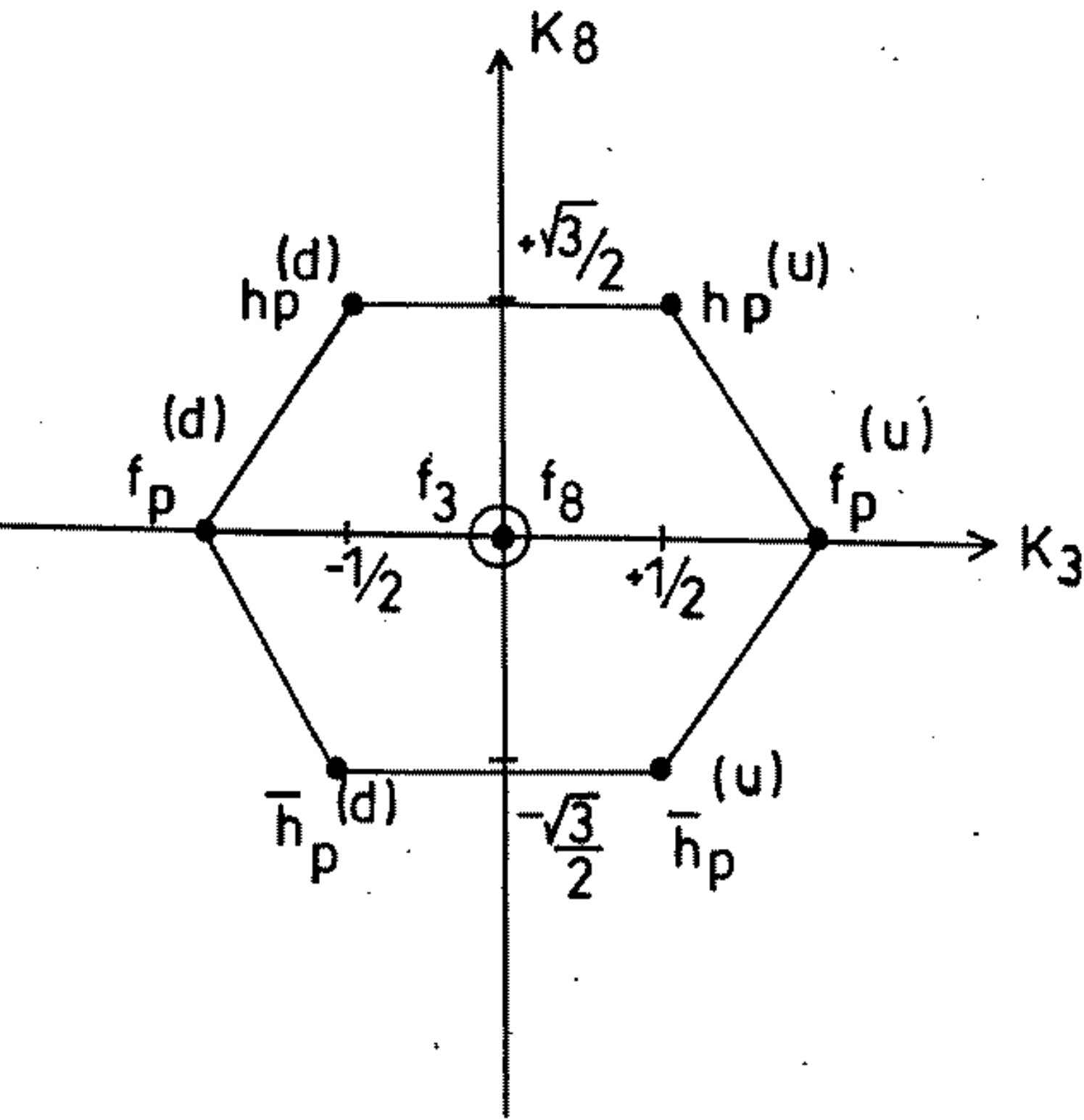
\item {Fig.3.} Weight diagram for gauge particles
\vfill
\eject
\end